\documentclass[12pt]{iopart}
\usepackage{epsfig}
\newcommand{\vev}[1]{\left\langle #1 \right\rangle}
\newcommand{\Sf}{S_{\mathrm{free}}}
\newcommand{\Si}{S_{\mathrm{int}}}
\newcommand{\fmslash}[1]{\hbox{$#1$\kern-0.5em\raise0.3ex\hbox{/}}}
\newcommand{\ld}[1]{\frac{\overrightarrow{\delta}}{\delta \bar{\psi}
      (#1)}}
\newcommand{\rd}[1]{\frac{\overleftarrow{\delta}}{\delta \psi (#1)}}
\newcommand{\Sp}{\mathrm{Sp}~}
\newcommand{\Op}{\mathcal{O}}
\newcommand{\ep}{\epsilon}
\newcommand{\zero}{\Bigg|_0}
\begin{document}

\title{On the construction of QED using ERG}\author{H Sonoda}
\address{Physics Department, Kobe University, Kobe 657-8501, Japan}
\ead{hsonoda@kobe-u.ac.jp}

\begin{abstract}
  It has been known for some time that a smooth momentum cutoff is
  compatible with local gauge symmetries.  In this paper we show
  concretely how to construct QED using the exact renormalization
  group (ERG).  First, we give a new derivation of the Ward identity
  for the Wilson action using the technique of composite operators.
  Second, parameterizing the theory by its asymptotic behavior for a
  large cutoff, we show how to fine-tune the parameters to satisfy the
  identity.  Third, we recast the identity as invariance of the Wilson
  action under a non-linear BRST transformation.
\end{abstract}

\submitto{\JPA}
\maketitle

\section{Introduction\label{intro}}

Ken Wilson introduced the exact renormalization group (ERG) to define
continuum limits in quantum field theory.\cite{wk} Using a theory with
a finite ultraviolet (UV) cutoff, we can construct a continuum limit
with the expense of keeping an infinite number of terms in the action.
The idea was subsequently applied to perturbation theory by Joe
Polchinski.\cite{joe} The purpose of this paper is to apply
Polchinski's version of ERG to construct QED perturbatively.

Before writing anything about gauge theories, we would like to clarify
our terminology.  The ``action'' we have mentioned above is a
functional of field variables.  The momenta of the fields are
restricted to the range below a UV cutoff $\Lambda$.  The interaction
vertices of the action are obtained by integrating out the
fluctuations of fields with momenta larger than $\Lambda$.  Hence, the
correlation functions calculated with this action contain the physics
of all momentum scales.  The action with a UV cutoff is often called a
``Wilson action'' in the literature, and we will follow this practice
here even though an ``action with a UV cutoff'' is perhaps more
descriptive.  The reason is that we wish to avoid confusion with the
``effective action'' with an infrared (IR) cutoff $\Lambda$ that
consists of the 1PI parts of the vertices of the Wilson action.  In
the limit $\Lambda \to 0$, the ``effective action'' with an IR cutoff
reduces to the standard effective action.  ERG determines the UV
cutoff dependence of the Wilson action or the IR cutoff dependence of
the ``effective action''.  Except for this introduction, we will
consider only the Wilson action in this paper.

The construction of gauge theories in the framework of ERG has been
discussed by various authors.  To the author's knowledge the first
person who applied ERG systematically to construct (non-abelian) gauge
theories is Becchi.\cite{becchi} (There is an even earlier work by
Warr \cite{warr}, who treated ERG mainly as a method of
regularization.)  Becchi showed how to construct a BRST invariant
Wilson action by introducing sources that generate BRST
transformations.  Alternatively gauge theories can be formulated for
the effective action with an IR cutoff.  This was initiated by
Ellwanger \cite{ellwanger} under the name of the ``modified'' Ward (or
Slavnov-Taylor) identities.

Whether we study a Wilson action with a UV cutoff or an effective
action with an IR cutoff, the form of the Ward identities depends on
the cutoff $\Lambda$ explicitly.  (This is the reason for the
adjective ``modified.'')  But ERG guarantees the $\Lambda$
independence of the identities: once the identities are satisfied for
some $\Lambda$, it is satisfied for any $\Lambda$.  It is this
property which simplifies the study of gauge symmetry within the ERG
framework.

The first explicit construction of QED using ERG was done in
\cite{bonini}.  They computed the effective action with $\Lambda = 0$
by solving ERG differential equations starting from a set of initial
conditions at a very large cutoff $\Lambda_0$.  They showed how to
fine tune the initial conditions so that the effective action with
$\Lambda = 0$ satisfies the Ward identities.  There, ERG is merely a
device for loop calculations, and they did not take advantage of the
simplicity of the Ward identities at large $\Lambda$.  Except for the
original work of Becchi, most later works seem to share the same
short-coming: the Ward identities are always checked at $\Lambda = 0$.

The main purpose of this paper is to construct the Wilson action of
QED by satisfying the Ward identities at large UV cutoff $\Lambda$.
As advocated by Becchi \cite{becchi}, the Ward identities get
simplified at large $\Lambda$, and to take advantage of this we need
to parameterize the Wilson action in terms of its asymptotic behavior
at large $\Lambda$.  Such parameterization scheme has been introduced
in refs.~\cite{integral,erg}, and we apply it here to QED.

We organize this paper as follows.  In sect.~2, we briefly review
Polchinski's ERG differential equation by applying it to QED.  In
sect.~3 we parameterize the Wilson action in terms of its behaviors at
large cutoff.  This is an application of the scheme explained in
\cite{erg}.  (The reader unfamiliar with perturbative applications of
ERG may find a casual reading of \cite{erg} helpful.)  In sects.~4 and
5 we give a new derivation of the Ward identities for the Wilson
action.  The result we obtain is equivalent to what has been derived
before by Becchi \cite{becchi} and Ellwanger \cite{ellwanger} and
others for YM theories. Our starting point is the Ward identities in
the continuum limit of QED.  A crucial observation, which has not been
emphasized or clearly understood in the previous literature on ERG, is
that the continuum limit can be constructed using a Wilson action with
any finite UV cutoff.  (This is derived for the real scalar theory in
appendix A.)  Hence, we can rewrite the Ward identities in the
continuum limit as the identities for the Wilson action.  We will end
up with an operator equation (\ref{id}), an equality between two
composite operators.  In sect.~6, we explain how we can satisfy
(\ref{id}) by fine tuning the parameters of the theory order by order
in loop expansions.  We give 1-loop calculations in sect.~7.  After
giving brief remarks about how to rewrite the Ward identity (\ref{id})
as BRST invariance in sect.~8, we conclude the paper in sect.~9.  Some
detailed calculations are given in appendix B.

Before closing we mention some relevant past works.  Besides
\cite{bonini}, construction of QED using ERG has been done 
by others, and we mention only a couple here.  In \cite{scalar,fw} the
background field method was applied to scalar QED.  More recently, a
manifestly gauge invariant formulation has been constructed not only
for QED \cite{arnone} but also for general YM theories.\cite{rosten}

Throughout the paper we use the euclidean metric.

\section{Polchinski's equation}

The Wilson action of QED is given as the sum of free and
interaction parts:
\begin{equation}
S (\Lambda) = \Sf (\Lambda) + \Si (\Lambda)
\end{equation}
where
\begin{eqnarray}
\Sf (\Lambda) &\equiv& - \frac{1}{2} \int_k A_\mu (k) A_\nu (-k)
\frac{1}{K(k/\Lambda)} \left( k^2 \delta_{\mu\nu} - \left(1 -
\frac{1}{\xi} \right) k_\mu k_\nu \right) \nonumber\\ && \, - \int_p
\frac{1}{K (p/\Lambda)} \bar{\psi} (-p) \left( \fmslash{p} + i m
\right) \psi (p)
\end{eqnarray}
We use a sign convention so that the weight of functional integration
is given by $\exp [S (\Lambda)]$ instead of the more usual $\exp [-
S(\Lambda)]$.  The momentum cutoff $\Lambda$ is introduced via a
positive cutoff function $K(x)$, which is $1$ for $x^2 < 1$ and decays
rapidly (e.g., exponentially) for $x^2 > 1$.  The free propagators are
given by
\begin{eqnarray}
\vev{A_\mu (k) A_\nu (-k)}_{\Sf (\Lambda)} &=& \frac{K(k/\Lambda)}{k^2} \left(
\delta_{\mu\nu} - (1-\xi) \frac{k_\mu k_\nu}{k^2} \right)\\
\vev{\psi (p) \bar{\psi} (-p)}_{\Sf (\Lambda)} &=&
\frac{K(p/\Lambda)}{\fmslash{p} + i m}
\end{eqnarray}
Throughout the paper we consider only the connected part of the
correlation functions.  For notational simplicity we omit the overall
factor of the delta function for momentum conservation.

We can construct the continuum limit using a Wilson action $S(\Lambda)$
with a finite $\Lambda$.  But this comes with a price: we must keep an
increasing number of terms in the Wilson action as we go to higher orders
in perturbation theory.  The $\Lambda$ dependence of the interaction
part $\Si (\Lambda)$ is given by Polchinski's ERG differential
equation \cite{joe}: 
\begin{eqnarray}
  &&- \Lambda \frac{\partial \Si (\Lambda)}{\partial \Lambda} = \int_k
  \frac{\Delta (k/\Lambda)}{k^2} \left( \delta_{\mu\nu} - (1-\xi)
    \frac{k_\mu k_\nu}{k^2} \right)\nonumber\\
&& \qquad\qquad\qquad
 \times \frac{1}{2} \left( \frac{\delta \Si (\Lambda)}{\delta A_\mu
      (k)} \frac{\delta \Si (\Lambda)}{\delta A_\nu (-k)} + \frac{\delta^2 \Si
      (\Lambda)}{\delta A_\mu (k) \delta A_\nu (-k)} \right)
  \nonumber\\ && 
\qquad\qquad - \int_p \Delta (p/\Lambda) \,\Sp \Bigg[
  \frac{1}{\fmslash{p}+im} \nonumber\\
&& \qquad \times \left\lbrace \ld{-p} \Si (\Lambda) \cdot \Si
    (\Lambda) \rd{p} + \ld{-p} \Si (\Lambda) \rd{p}\right\rbrace
  \Bigg]
\label{joe}
\end{eqnarray}
where
\begin{equation}
\Delta \left(\frac{p}{\Lambda}\right) \equiv \Lambda
\frac{\partial}{\partial \Lambda} K \left(\frac{p}{\Lambda}\right)
\end{equation}
is a non-negative function which is zero for $p^2 < \Lambda^2$, has an
appreciative value only for $p^2 \sim \Lambda^2$, and decays rapidly
as $p^2$ grows.  The minus sign in front of the second integral of
(\ref{joe}) is due to the Fermi statistics.

It is common to think that $S(\Lambda)$ describes only the physics of
low momentum $p < \Lambda$ correctly.  On a close inspection, however,
we find that $S(\Lambda)$ contains the physics of all momentum scales.
To be precise, the renormalized two-point functions are given by
\begin{eqnarray}
\vev{A_\mu (-k) A_\nu (k)} &=& \frac{1 - K(k/\Lambda)^{-1}}{k^2}
\left( \delta_{\mu\nu} - (1-\xi) \frac{k_\mu k_\nu}{k^2} \right)
\nonumber\\ && \qquad + \frac{1}{K(k/\Lambda)^2} \vev{A_\mu (-k) A_\nu
(k)}_{S (\Lambda)} \label{aa}\\
\vev{\psi (p) \bar{\psi} (-p)} &=& \frac{1 -
  K(p/\Lambda)^{-1}}{\fmslash{p} + i m} + \frac{1}{K(p/\Lambda)^2}
  \vev{\psi (p) \bar{\psi} (-p)}_{S (\Lambda)}\label{psipsibar}
\end{eqnarray}
where $\vev{\cdots}_S$ is the correlation function calculated with
$S$.  The renormalized higher point functions are more simply given by
\begin{eqnarray}
  &&\vev{A_{\mu_1} (k_1) \cdots A_{\mu_M} (k_M) \psi (p_1) \cdots \psi
    (p_N) \bar{\psi} (-q_1) \cdots \bar{\psi} (-q_N)}\nonumber\\ && =
  \prod_{i=1}^M \frac{1}{K(k_i/\Lambda)} \prod_{i=1}^N \frac{1}{K
    (p_i/\Lambda) K(q_i/\Lambda)} \nonumber\\
  && \qquad \cdot \vev{A_{\mu_1} (k_1) \cdots
    A_{\mu_M} (k_M) \psi (p_1) \cdots \psi (p_N) \bar{\psi} (-q_1)
    \cdots \bar{\psi} (-q_N)}_{S(\Lambda)}\label{higher}
\end{eqnarray}
To make sense of the division by $K (p/\Lambda)$, we assume that
$K(p/\Lambda)$ is strictly positive and that it vanishes only in the
limit $p^2 \to \infty$.  As long as the Wilson action $S(\Lambda)$
satisfies (\ref{joe}), the right-hand sides of
(\ref{aa},\ref{psipsibar},\ref{higher}) are independent of the cutoff
$\Lambda$.  Since the results (\ref{aa},\ref{psipsibar},\ref{higher})
may not be widely known, we derive analogous results for the real
scalar field theory in appendix A.

\section{Parameterization}

Let us next discuss how to parameterize the solutions of (\ref{joe}).
We are only interested in those solutions corresponding to
renormalized theories.  (See \cite{integral, erg} for a similar
analysis for the $\phi^4$ theory.)  We evaluate the functional derivatives
of $S(\Lambda)$ at vanishing fields.  The asymptotic behaviors at
large $\Lambda$ are obtained by expansions in powers of the electron
mass and external momenta:
\numparts
\begin{eqnarray}
&&\frac{\delta^2 \Si}{\delta A_\mu (k) \delta A_\nu (-k)}\zero =
\delta_{\mu\nu} \Bigg\lbrace a_2 \left(\ln \Lambda/\mu\right)
\Lambda^2 \nonumber\\
&&\quad + b_2 \left( \ln \Lambda/\mu \right) m^2 + c_2
\left(\ln \Lambda/\mu\right) k^2 \Bigg\rbrace + d_2 \left(\ln
\Lambda/\mu \right) k_\mu k_\nu + \cdots \label{asymp-aa}\\
&&\ld{-p} \Si \rd{p}\zero = a_f \left( \ln \Lambda/\mu \right)
\fmslash{p} + b_f \left( \ln \Lambda/\mu \right) i m + \cdots\\
&&\ld{-p-k} \frac{\delta S}{\delta A_\mu (k)} \rd{p}\zero = a_3 \left( \ln
\Lambda/\mu \right) \gamma_\mu + \cdots \label{asymp-vertex}\\
&&\frac{\delta^4 S}{\delta A_\alpha (k_1) \delta A_\beta (k_2) \delta
  A_\gamma (k_3) \delta A_\delta (k_4)}\zero \nonumber\\
&& \qquad\qquad = a_4 \left( \ln
\Lambda/\mu \right) \left( \delta_{\alpha\beta} \delta_{\gamma\delta}
  + \delta_{\alpha\gamma} \delta_{\beta\delta} + \delta_{\alpha
  \delta} \delta_{\beta \gamma} \right) + \cdots \label{asymp-aaaa}
\end{eqnarray}
\endnumparts
where an arbitrary momentum scale $\mu > 0$ is introduced to make
$\frac{\Lambda}{\mu}$ dimensionless.  The vertical line with subscript
$0$ is a reminder that the derivative is evaluated at vanishing
fields.  (The extraction of the asymptotic parts is similar to the
$\mathcal{T}$ operation in \cite{becchi}.)

For the renormalizable theories, the dotted parts are suppressed by
negative powers of $\Lambda$, hence vanishing in the limit
$\Lambda/\mu \to \infty$.  The higher order derivatives have negative
dimensions, and vanish also in the limit $\Lambda/\mu \to \infty$.

In solving (\ref{joe}), the seven functions
\begin{equation}
\begin{array}{c}
b_2 (\ln \Lambda/\mu), c_2(\ln \Lambda/\mu) , d_2(\ln
\Lambda/\mu),\\ a_f(\ln \Lambda/\mu) , b_f(\ln
\Lambda/\mu) , a_3(\ln \Lambda/\mu) ,a_4 (\ln \Lambda/\mu)
\end{array}
\end{equation}
are ambiguous by additive constants.  (Note that the quadratically
divergent $\Lambda^2 a_2$ has no such
ambiguity.)  Therefore, in order to specify a unique solution of
(\ref{joe}), we must introduce seven conditions.  To begin, we can
adopt the following three normalization conditions:
\begin{equation}
c_2 (0) = a_f (0) = b_f (0) = 0
\end{equation}
($c_2 (0)$ normalizes the gauge field, $a_f (0)$ the spinor fields,
and $b_f (0)$ the mass parameter $m$.)  It is one of the main purposes
of this paper to show how the Ward identities determine the remaining
four constants
\begin{equation}
b_2 (0), d_2 (0), a_3 (0), a_4 (0) \label{four}
\end{equation}
in terms of the elementary charge $e$, which itself is introduced
through the Ward identities.

\section{The Ward identities}

Let us recall the renormalized Ward identities in QED.  They have two
parts.  The first part is the identity for the two-point function of
the gauge field:
\begin{equation}
\frac{1}{\xi} k_\mu \vev{A_\mu (-k) A_\nu (k)} = \frac{k_\nu}{k^2}
\label{ward-1} 
\end{equation}
The second part is the Ward identities for the higher-point functions:
\begin{eqnarray}
&&\frac{1}{\xi} k_\mu \vev{A_\mu (-k) A_{\mu_1} (k_1) \cdots A_{\mu_M}
  (k_M) \psi (p_1) \cdots \psi (p_N) \bar{\psi} (-q_1) \cdots
  \bar{\psi} (-q_N)} \nonumber\\ &&= \frac{e}{k^2} \sum_{i=1}^N \left[
  \vev{A_{\mu_1} (k_1) \cdots \psi (p_i - k) \cdots} - \vev{A_{\mu_1}
  (k_1) \cdots \bar{\psi} (-q_i-k) \cdots} \right] \label{ward-2}
\end{eqnarray}
where $e$ is interpreted as the elementary charge defined at scale
$\mu$.  The Ward identities can be also written as the conservation
law:
\begin{eqnarray}
&&k_\mu \vev{J_\mu (-k) A_{\mu_1} (k_1) \cdots A_{\mu_M}
  (k_M) \psi (p_1) \cdots \psi (p_N) \bar{\psi} (-q_1) \cdots
  \bar{\psi} (-q_N)} \nonumber\\ &&= e \sum_{i=1}^N \left[
  \vev{A_{\mu_1} (k_1) \cdots \psi (p_i - k) \cdots} - \vev{A_{\mu_1}
  (k_1) \cdots \bar{\psi} (-q_i-k) \cdots} \right] \label{cvc}
\end{eqnarray}
where $J_\mu$ is the charge current.  The goal of this section is to
find a concrete expression of $J_\mu$ in terms of the interaction
action $\Si (\Lambda)$.

We first examine (\ref{ward-1}).  Using (\ref{aa}), we can
rewrite (\ref{ward-1}) as
\begin{equation}
\frac{1}{\xi} k_\mu \vev{A_\mu (-k) A_\nu (k)}_{S(\Lambda)} = K(k/\Lambda)
\frac{k_\nu}{k^2} \label{ward-1a}
\end{equation}
Since
\begin{eqnarray}
&&\vev{A_\mu (-k) A_\nu (k)}_{S(\Lambda)} \nonumber\\
&& \quad= \frac{K(k/\Lambda)}{k^2}
\left( \delta_{\mu\alpha} - (1-\xi) \frac{k_\mu k_\alpha}{k^2} \right)
\cdot \left( \delta_{\alpha\nu} +
\vev{\frac{\delta \Si}{\delta A_\alpha (k)} A_\nu (k)}_{S(\Lambda)}
\right)\label{firstrelation}
\end{eqnarray}
(see appendix A for an analogous result for the real scalar theory), we
can rewrite (\ref{ward-1a}) as
\begin{equation}
k_\mu \vev{\frac{\delta \Si}{\delta
    A_\nu (k)} A_\nu (k)}_{S(\Lambda)} = 0 \label{ward0}
\end{equation}

Next, we examine (\ref{ward-2}).  Using (\ref{higher}) and
\begin{eqnarray}
&&\vev{A_\mu (-k) A_{\mu_1} (k_1) \cdots \psi (p_1) \cdots \bar{\psi}
  (-q_1) \cdots}_{S(\Lambda)} \nonumber\\ && = \frac{K(k/\Lambda)}{k^2}
  \left( \delta_{\mu\alpha} - (1-\xi) \frac{k_\mu k_\alpha}{k^2}
  \right)\nonumber\\
&&\qquad\cdot \vev{\frac{\delta \Si}{\delta A_\alpha (k)}
  A_{\mu_1} (k_1) \cdots \psi (p_1) \cdots \bar{\psi} (-q_1)
  \cdots}_{S(\Lambda)}\label{secondrelation}
\end{eqnarray}
(again see appendix A), we can rewrite
(\ref{ward-2}) as
\begin{eqnarray}
&& k_\mu \vev{\frac{\delta \Si}{\delta A_\mu (k)} A_{\mu_1} (k_1)
    \cdots A_{\mu_M} (k_M) \psi (p_1) \cdots \psi (p_N) \bar{\psi}
    (-q_1) \cdots \bar{\psi} (-q_N) }_{S(\Lambda)}\nonumber\\ &&
    \qquad \times \frac{1}{\prod_{i=1}^M K(k_i/\Lambda) \prod_{j=1}^N
    K(p_i/\Lambda) K(q_i/\Lambda)} \nonumber\\ && = e \sum_{n=1}^N
    \Bigg[ \vev{ A_{\mu_1} (k_1) \cdots \psi (p_n - k) \cdots} -
    \vev{A_{\mu_1} (k_1) \cdots \bar{\psi}(-q_n-k) \cdots}
    \Bigg]\label{ward}
\end{eqnarray}

To summarize so far, we can replace (\ref{ward-1},\ref{ward-2}) by
(\ref{ward0},\ref{ward}).  Note that (\ref{ward0}) is a special case
$M=1$, $N=0$ of (\ref{ward}).  Hence, (\ref{ward}) for arbitrary $M,
N$ is equivalent to (\ref{ward-1},\ref{ward-2}).

We now define a vector current by
\begin{equation}
J_\mu (-k) \equiv \frac{\delta \Si (\Lambda)}{\delta A_\mu (k)} \label{jmu}
\end{equation}
This is a composite operator since its correlation functions
\begin{eqnarray}
&&\vev{J_\mu (-k) A_{\mu_1} (k_1) \cdots \psi (p_1) \cdots
    \bar{\psi}(-q_1) \cdots}\nonumber\\ && \equiv \vev{J_\mu (-k)
    A_{\mu_1} (k_1) \cdots \psi (p_1) \cdots \bar{\psi}(-q_1)
    \cdots}_{S(\Lambda)}\nonumber\\ && \qquad \times
    \frac{1}{\prod_{i=1}^M K (k_i/\Lambda) \prod_{j=1}^N
    K(p_j/\Lambda) K (q_j/\Lambda)} \label{Jcorr}
\end{eqnarray}
are independent of $\Lambda$.  (See Appendix A for an explanation on
composite operators.)  To show this, we differentiate
(\ref{joe}) with respect to $A_\mu (k)$ to obtain
\begin{equation}
- \Lambda \frac{\partial}{\partial \Lambda} J_\mu (-k) = \mathcal{D}
  \cdot J_\mu (-k) \label{operator}
\end{equation}
where $\mathcal{D}$ is a linear differential operator defined by
\begin{eqnarray}
&& \mathcal{D} \cdot \mathcal{F} \equiv \int_l \frac{\Delta (l/\Lambda)}{l^2}
\left( \delta_{\mu\nu} - (1-\xi) \frac{l_\mu l_\nu}{l^2} \right)
\nonumber\\ &&\qquad\qquad\qquad\qquad \cdot \left( \frac{\delta
\Si}{\delta A_\mu (-l)} \frac{\delta \mathcal{F}}{\delta A_\nu (l)} +
\frac{1}{2} \frac{\delta^2 \mathcal{F} }{\delta A_\mu (l) \delta A_\nu (-l)}
\right) \nonumber\\ && \qquad\quad - \int_p \Delta (p/\Lambda) \, \Sp
\frac{1}{\fmslash{p} + i m} \Bigg\lbrace \ld{-p} \Si \cdot \mathcal{F}
\rd{p}\nonumber\\ &&\qquad\qquad\qquad + \ld{-p} \mathcal{F} \cdot \Si \rd{p}
+ \ld{-p} \mathcal{F} \rd{p} \Bigg\rbrace
\end{eqnarray}
acting on any functional $\mathcal{F}$.  (\ref{operator}) guarantees
that the correlation functions (\ref{Jcorr}) are independent of
$\Lambda$.  Thus, from (\ref{ward}--\ref{Jcorr}), we can
rewrite the Ward identities as (\ref{cvc}) where $J_\mu$ is defined by
(\ref{jmu}).  (For a discussion of composite operators in $\phi^4$
theory, see \cite{erg}.)

\section{Ward identity as an operator identity \label{opid}}

In this section we wish to rewrite the Ward identities
(\ref{cvc}) as a single operator equation.  For this purpose we 
construct a composite operator $\Phi (-k)$ whose correlation functions
give the right-hand side of (\ref{cvc}):
\begin{eqnarray}
&&\vev{\Phi (-k) A_{\mu_1} (k_1) \cdots A_{\mu_M} (k_M) \psi (p_1)
  \cdots \psi (p_N) \bar{\psi} (-q_1) \cdots \bar{\psi} (-q_N)}
  \nonumber\\ 
&&  = e \sum_{i=1}^N \left( \vev{A_{\mu_1} (k_1) \cdots \psi (p_i-k)
  \cdots} - \vev{A_{\mu_1} (k_1) \cdots \bar{\psi} (-q_i - k) \cdots}
  \right) \label{Phicorr}
\end{eqnarray} 
Once we construct $\Phi (-k)$, the Ward identities (\ref{cvc}) are
equivalent to the single operator equation:
\begin{equation}
k_\mu J_\mu (-k) = \Phi (-k) \label{id}
\end{equation}
The construction of $\Phi (-k)$ is essential, but it involves a fair
amount of technicality.  The derivation will be given in appendix B.
There, we show that $\Phi (-k)$ is given by
\begin{eqnarray}
\Phi (-k) &\equiv& \int_p \Sp U (-p-k,p) \nonumber\\ &&\qquad \cdot
\left\lbrace \ld{-p} S \cdot S \rd{p+k} + \ld{-p} S \rd{p+k}
\right\rbrace \nonumber\\ && + e \int_p \Bigg[ - S \rd{p}
\frac{K(p/\Lambda)}{K\left((p-k)/\Lambda\right)} \psi (p-k)\nonumber\\
&&\qquad\qquad + \frac{K(p/\Lambda)}{K\left((p+k)/\Lambda\right)} \bar{\psi}
(-p-k) \ld{-p} S \Bigg] \label{Phi}
\end{eqnarray}
where the matrix $U$ is defined by
\begin{eqnarray}
&&U (-p-k, p) \nonumber\\ && \equiv e \left[ K \left( (p+k)/\Lambda
\right) \frac{1 - K(p/\Lambda)}{\fmslash{p} + i m} - \frac{1 - K
\left( (p+k)/\Lambda \right)}{\fmslash{p} + \fmslash{k} + i m} K
(p/\Lambda) \right] \label{U}
\end{eqnarray}
$\Phi$ is a composite operator, satisfying
\begin{equation}
- \Lambda \frac{\partial}{\partial \Lambda} \Phi (-k) = \mathcal{D}
  \cdot \Phi (-k)
\end{equation}
Hence, if (\ref{id}) is valid asymptotically for large $\Lambda \gg
\mu$, then it is valid for any $\Lambda$.  As explained in sect.~1,
this is the most important feature of the Ward identity in ERG.  A
Ward identity similar to (\ref{id}) has been obtained for the scalar
QED in equation (7) of \cite{fw}, using the background field formalism.

\section{Fine tuning}

We now show how to satisfy the Ward identity (\ref{id}) for $\Lambda
\gg \mu$ by fixing the four constants (\ref{four}).

The asymptotic behavior of $k_\mu J_\mu$ is easily obtained from the
asymptotic behavior of the Wilson action, which is given by
(\ref{asymp-aa},\ref{asymp-vertex},\ref{asymp-aaaa}):
\numparts
\begin{eqnarray}
&&\frac{\delta k_\mu J_\mu (-k)}{\delta A_\nu (-k)}\zero = k_\nu
\Bigg\lbrace a_2 (\ln \Lambda/\mu) \Lambda^2 \nonumber\\
&&\qquad\qquad\qquad + b_2 (\ln \Lambda/\mu) m^2 + \left(c_2 +
d_2\right) (\ln \Lambda/\mu) k^2 \Bigg\rbrace + \cdots\\ && \ld{-p-k}
k_\mu J_\mu (-k) \rd{p}\zero = a_3 (\ln \Lambda/\mu) \fmslash{k} + \cdots\\
&&\frac{\delta^3 k_\mu J_\mu (-k)}{\delta A_\alpha (k_1) \delta
A_\beta (k_2) \delta A_\gamma (k_3)}\zero \nonumber\\ &&\qquad\qquad\qquad
= a_4 (\ln \Lambda/\mu) \left( k_{\alpha} \delta_{\beta\gamma} +
k_\beta \delta_{\alpha\gamma} + k_\gamma \delta_{\alpha\beta} \right)
+ \cdots
\end{eqnarray}
\endnumparts

The composite operator $\Phi (-k)$ is a dimension $4$ scalar operator
of momentum $-k$, odd under charge conjugation, and vanishing at
$k=0$.  Hence, its asymptotic behavior has the same form as above:
\numparts
\begin{eqnarray}
&&\frac{\delta \Phi}{\delta A_\nu (- k)}\Bigg|_0 \nonumber\\
&& \quad = k_\nu \Bigg\lbrace \bar{a}_2 (\ln \Lambda/\mu) \Lambda^2
 + \bar{b}_2 (\ln \Lambda/\mu) m^2 + \bar{d}_2 (\ln
\Lambda/\mu) k^2 \Bigg\rbrace + \cdots\\
&& \ld{-p-k} \Phi (-k) \rd{p}\Bigg|_0 = \bar{a}_3 (\ln \Lambda/\mu)
 \fmslash{k} + \cdots\\
&&\frac{\delta^3 \Phi (-k)}{\delta A_\alpha
  (k_1) \delta A_\beta (k_2) \delta A_\gamma (k_3)}\Bigg|_0\nonumber\\
&& \qquad\qquad = \bar{a}_4 (\ln \Lambda/\mu) \left( k_{\alpha}
  \delta_{\beta\gamma} + k_\beta \delta_{\alpha\gamma} + k_\gamma
  \delta_{\alpha\beta} \right) + \cdots
\end{eqnarray}
\endnumparts
The values of the four dimensionless functions $\bar{b}_2$,
$\bar{d}_2$, $\bar{a}_3$, $\bar{a}_4$ at $\ln \Lambda/\mu = 0$
determine $\Phi (-k)$ uniquely.  Therefore, the operator equation
(\ref{id}) is equivalent to the following four equations:
\begin{equation}
\begin{array}{c@{~=~}c}
b_2 (0) & \bar{b}_2 (0)\\
d_2 (0) & \bar{d}_2 (0)\\
a_3 (0) & \bar{a}_3 (0)\\
a_4 (0) & \bar{a}_4 (0)
\end{array}\label{recursion}
\end{equation}
where we have used the convention $c_2 (0) = 0$.

Now, $\Phi$ is defined by (\ref{Phi}, \ref{U}) in terms of the Wilson action
$S (\Lambda)$.  Therefore, we obtain
\numparts
\begin{eqnarray}
&&\frac{\delta \Phi}{\delta A_\nu (- k)}\Bigg|_0 = \int_q \Sp U (-q-k,
 q) \ld{-q} \frac{\delta S}{\delta A_\nu (-k)} \rd{q+k}\zero
 \label{Phi-1}\\ && \ld{-p-k} \Phi (-k) \rd{p} \zero = e \left(1 - a_f
 (\ln \Lambda/\mu) \right) \fmslash{k}\nonumber\\ &&\quad + \int_q
 \ld{-p-k} \Sp \left\lbrace U(-q-k,q) \ld{-q} S \rd{q+k} \right\rbrace
 \rd{p} \zero \label{Phi-2}\\ &&\frac{\delta^3 \Phi (-k)}{\delta
 A_\alpha (k_1) \delta A_\beta (k_2) \delta A_\gamma
 (k_3)}\Bigg|_0\nonumber\\ && \, = \int_q \Sp U (-q-k,q) \ld{-q}
 \frac{\delta^3 S}{\delta A_\alpha (k_1) \delta A_\beta (k_2) \delta
 A_\gamma (k_3)} \rd{q+k} \zero\label{Phi-3}
\end{eqnarray}
\endnumparts
Expanding the right-hand sides in powers of the electron mass and
external momenta, we can obtain $\bar{b}_2$, $\bar{d}_2$, $\bar{a}_3$,
$\bar{a}_4$.  On the right-hand sides one extra loop is given
explicitly by the integral over $q$, and it is sufficient to know $S
(\Lambda)$ only up to $(l-1)$-loop level to determine the four
constants $\bar{b}_2 (0), \bar{d}_2 (0), \bar{a}_3 (0), \bar{a}_4 (0)$
at $l$-loop level.  (Note $a_f (0) =0$ by convention.)  In other
words, with $b_2 (0), d_2 (0), a_3 (0), a_4 (0)$ determined up to
$(l-1)$-loop level, we can determine the $l$-loop values of $b_2 (0),
d_2 (0), a_3 (0), a_4 (0)$ by demanding (\ref{recursion}).  We can
thus construct QED using the Ward identities.  This recursive nature
of the Ward identities has been observed by Becchi \cite{becchi} and
Morris and D'Attanasio \cite{md} among others.  We have taken
advantage of it using our particular parameterization of the theory.

\section{1-loop calculations}

Using the results of the previous section, let us do 1-loop
calculations.  We start from the tree-level values:
\begin{equation}
 b_2 (0) = d_2 (0) = a_4 (0) = 0,\quad a_3 (0) = e
\end{equation}

At 1-loop, (\ref{Phi-1}) gives
\begin{equation}
\frac{\delta \Phi}{\delta A_\nu (-k)}\zero = e^2 \int_q \Sp
  \gamma_\nu U (-q-k,q)
\end{equation}
Changing the integration variable as
\begin{equation}
q \longrightarrow \Lambda q \label{change}
\end{equation}
we obtain
\begin{eqnarray}
&& \frac{\delta \Phi}{\delta A_\nu (-k)}\zero = e^2 \Lambda^2 \int_q
  \Sp \gamma_\nu \nonumber\\ &&\qquad\qquad \cdot \left\lbrace K\left( q +
  \frac{k}{\Lambda} \right) \frac{1 - K (q)}{\fmslash{q} + i
  m/\Lambda} - \frac{1 - K\left( q +
  \frac{k}{\Lambda}\right)}{\fmslash{q} + \fmslash{k}/\Lambda + i
  m/\Lambda } K(q) \right\rbrace\nonumber\\ &&\quad\simeq e^2 k_\nu \left[ - 2
  \Lambda^2 \int_q \frac{1}{q^2} \Delta (q) \left( 1 - K(q) \right) +
  \left( m^2 + \frac{k^2}{3} \right) \int_q \frac{\Delta (q)}{(q^2)^2}
  \right]
\end{eqnarray}
where we have expanded the integral in powers of $m/\Lambda$
and $k_\mu/\Lambda$.  Using
\begin{equation}
\int_q \frac{ \Delta (q) K(q)^n}{(q^2)^2}  = \frac{1}{n+1}
\int \frac{d^4 q}{(2 \pi)^4} \frac{- 2}{q^2} \frac{d}{d q^2} K( q )^{n+1} =
\frac{1}{(4 \pi)^2} \frac{2}{n+1}
\end{equation}
(independent of the choice of $K$), we obtain
\begin{equation}
b_2 (0) = \frac{2 e^2}{(4 \pi)^2},\quad
d_2 (0) = \frac{2 e^2}{3 (4 \pi)^2}
\end{equation}

Next we consider (\ref{Phi-2}) at 1-loop:
\begin{eqnarray}
&&\ld{-p-k} \Phi (-k) \rd{p} \zero - e (1 - a_f (\ln \Lambda/\mu))
\fmslash{k} \nonumber\\
&& = - e^2 \int_q \gamma_\mu U (-q-k,q) \gamma_\nu\nonumber\\
&& \qquad \cdot \frac{1 - K ((q-p)/\Lambda)}{(q-p)^2} \left(
\delta_{\mu\nu} - (1-\xi) \frac{(q-p)_\mu (q-p)_\nu}{(q-p)^2}
\right)\nonumber\\
&& \simeq - e^3 \fmslash{k} \int_q \frac{1}{(q^2)^2} \left\lbrace \xi
K(q) \left(1 - K(q) \right)^2 + \frac{3 - \xi}{4} \left(1 -
K(q)\right) \Delta (q) \right\rbrace
\end{eqnarray}
where we have used the change of variables (\ref{change}), and
expanded the integral in powers of $k_\mu/\Lambda$, $p_\mu/\Lambda$.
Hence,
\begin{equation}
a_3 (0) - e = - e^3 \left( \xi \int_q \frac{1}{(q^2)^2} K(q) (1 -
K(q))^2 + \frac{3 - \xi}{4 (4\pi)^2} \right)
\end{equation}

Finally, we consider (\ref{Phi-3}):
\begin{eqnarray}
&&\frac{\delta^3 \Phi (-k)}{\delta
 A_\alpha (k_1) \delta A_\beta (k_2) \delta A_\gamma
 (k_3)}\Bigg|_0\nonumber\\
&& = e^3 \int_q \Sp U (-q-k,q) \nonumber\\
&&\quad \cdot \Bigg\lbrace \gamma_\gamma \frac{1 - K \left( (q-k_3)/\Lambda
 \right)}{\fmslash{q} - \fmslash{k}_3 + i m} \gamma_\beta 
\frac{1 - K\left( (q-k_2-k_3)/\Lambda \right)}{\fmslash{q} -
 \fmslash{k}_2 - \fmslash{k}_3 + i m} \gamma_\alpha\nonumber\\
&& \qquad\qquad + (\textrm{5 permutations}) \quad\Bigg\rbrace\nonumber\\
&& \simeq 2 e^4 \left( k_\alpha \delta_{\beta\gamma} + k_\beta
  \delta_{\gamma \alpha} + k_\gamma \delta_{\alpha\beta} \right)
\int_q \frac{\Delta (q/\Lambda)}{(q^2)^2} \left(1 - K(q/\Lambda)\right)^2
\end{eqnarray}
Hence, we obtain
\begin{equation}
a_4 (0) = \frac{4}{3} \frac{e^4}{(4 \pi)^2}
\end{equation}

\section{BRST invariance}

To elucidate the gauge structure of QED further, we introduce ghost
and antighost fields $c, \bar{c}$.  These fields are non-interacting,
and we define the total Wilson action by
\begin{equation}
\bar{S} (\Lambda) \equiv S (\Lambda) - \int_k \bar{c} (-k) c(k)
\frac{k^2}{K (k/\Lambda)}
\end{equation}
We now define the BRST transformation as follows:
\numparts
\begin{eqnarray}
\delta_\ep A_\mu (k) &=& k_\mu \ep c (k)\\
\delta_\ep c (k) &=& 0\\
\delta_\ep \bar{c} (-k) &=& - \frac{1}{\xi} k_\mu A_\mu (-k) \ep\\
\delta_\ep \psi (p) &=& \int_k \ep c(k) \Bigg( e \frac{K(p/\Lambda)}{K\left(
  (p-k)/\Lambda \right)} \psi (p-k) \nonumber\\
&& \qquad\qquad - U(-p,p-k) \ld{-p+k} \bar{S} \Bigg)\\
\delta_\ep \bar{\psi} (-p) &=& - \int_k \ep c(k)
e \frac{K(p/\Lambda)}{K\left( (p+k)/\Lambda \right)} \bar{\psi} (-p-k)
\end{eqnarray}
\endnumparts
where $\ep$ is an arbitrary Grassmann variable.  The transformation of
$\psi$ is highly non-linear due to the second term.  The Ward identity
(\ref{id}) is equivalent to the following BRST invariance:
\begin{equation}
\delta_\ep \bar{S} + \int_k \ep c (k) \int_p \Sp
  U(-p-k,p) \ld{-p} \bar{S} \rd{p+k} 
 = 0 \label{brst}
\end{equation}
The first term is the change of the Wilson action $\bar{S}$ under the
transformation.  The second term is the change of the fermionic
integration measure.  Note that the non-vanishing $\delta_{\epsilon}
\bar{S}$ implies the non-invariance of the action under the BRST
transformation.  For example, the non-invariance due to the photon
mass term is cancelled precisely by the jacobian.

Due to the electron number conservation, the transformation of $\psi,
\bar{\psi}$ under BRST is not unique.  Alternatively, we can define
the BRST transformation which is symmetric between $\psi$ and
$\bar{\psi}$: \numparts
\begin{eqnarray}
\delta_\ep \psi (p) &=& e K(p/\Lambda) \int_k \ep c(k) \Psi (p-k)\\
\delta_\ep \bar{\psi} (-p) &=& - e K(p/\Lambda) \int_k \ep c(k)
\bar{\Psi} (-p-k)
\end{eqnarray}
\endnumparts
where
\numparts
\begin{eqnarray}
\Psi (p) &\equiv& \psi (p) + \frac{1 - K(p/\Lambda)}{\fmslash{p} + i m}
\frac{\overrightarrow{\delta}}{\delta \bar{\psi} (-p)} \Si\\
\bar{\Psi} (-p) &\equiv& \bar{\psi} (-p) + \Si
\frac{\overleftarrow{\delta}}{\delta \psi (p)} \frac{1 -
  K(p/\Lambda)}{\fmslash{p} + i m}
\end{eqnarray}
\endnumparts
are composite operators.  The transformations of $A_\mu, c, \bar{c}$
are the same as before.  The BRST invariance is now given as
\begin{equation}
\delta_\ep \bar{S} - \int_k \Sp \left[
 \delta_\ep \psi (p) \frac{\overleftarrow{\delta}}{\delta \psi (p)} +
  \frac{\overrightarrow{\delta}}{\delta \bar{\psi} (-p)} \delta_\ep
  \bar{\psi} (-p) \right] = 0
\end{equation}

The first form of BRST invariance has been extended further using the
antifield formalism in the joint work with Igarashi and Itoh.\cite{IIS}

\section{Conclusions}

For some time the compatibility of a smooth cutoff with gauge
invariance has been established by the previous works
(\cite{becchi},\cite{ellwanger},\cite{bonini},\cite{scalar}--\cite{rosten}
and references therein) on ERG.  Nevertheless, the contrary viewpoint
is still prevalent among the practitioners of quantum field theory.
A main task of this paper was to introduce a concrete application of
ERG to convince the reader of the compatibility.  In particular we
have shown how to construct QED using ERG.  The Wilson action with a
finite momentum cutoff has exactly the same gauge invariance as the
continuum limit.  Nothing is compromised.  The crucial observation for
our construction is that with a Wilson action we can compute the
correlation functions in the continuum limit.  (See
(\ref{aa}--\ref{higher}) and appendix A.)

As we have mentioned in sect.~\ref{intro}, it is some time since the
ground work was laid down for incorporation of gauge symmetries in the
ERG formalism.  As far as formalism goes, Becchi's results in
\cite{becchi} are seminal, if not final.  A merit of the present paper
is in the use of a particular parameterization of the theory, based on
the asymptotic behavior of the Wilson action at a very large cutoff.
The Ward identities of the Wilson action simplify for a large cutoff,
and we were able to take advantage of it using our parameterization
scheme.  In a future publication we wish to apply the same scheme to
YM theories.

\ack

I thank Profs. Y.~Igarashi and K.~Itoh for the observation that the
Ward identity (\ref{id}) can be interpreted as the BRST invariance
(\ref{brst}).  Without their help I would have missed the BRST
invariance.  I also thank Prof. K.~Higashijima for a comment which
helped me to simplify the presentation of this paper.  This work was
partially supported by the Grant-In-Aid for Scientific Research from
the Ministry of Education, Culture, Sports, Science, and Technology,
Japan (\#14340077).

\appendix

\section{The cutoff dependence of the correlation functions}

The purpose of this appendix is to explain and derive
(\ref{aa},\ref{psipsibar},\ref{higher}) and
(\ref{firstrelation},\ref{secondrelation}) which play important roles
in the main text.  For simplicity, we derive the analogous results for
the simpler theory of a real scalar field, which is defined by the
Wilson action:
\begin{equation}
S (\Lambda) = \Sf (\Lambda) + \Si (\Lambda)
\end{equation}
where
\begin{eqnarray}
\Sf (\Lambda) &=& - \frac{1}{2} \int_p \phi (p) \phi (-p) \frac{p^2 +
  m^2}{K(p/\Lambda)}\\
\Si (\Lambda) &=& \sum_{n=1}^\infty \frac{1}{(2n)!}
\int_{p_1,\cdots,p_{2n}} \mathcal{V}_{2n} (\Lambda; p_1,\cdots,p_{2n})
\phi (p_1) \cdots \phi (p_{2n})
\end{eqnarray}
The interaction part satisfies the following Polchinski differential
equation:
\begin{equation}
- \Lambda \frac{\partial \Si (\Lambda)}{\partial \Lambda}
= \int_p \frac{\Delta (p/\Lambda)}{p^2 + m^2} \frac{1}{2} \left\lbrace
  \frac{\delta \Si}{\delta \phi (p)} \frac{\delta \Si}{\delta \phi
    (-p)} + \frac{\delta^2 \Si}{\delta \phi (p) \delta \phi (-p)}
\right\rbrace \label{scalar}
\end{equation}
The correlation functions are computed perturbatively using the
propagator 
\begin{equation}
\frac{K(p/\Lambda)}{p^2 + m^2}
\end{equation}
and the interaction vertices $\mathcal{V}_{2n}$.  We denote the
connected part of the $n$-point correlation function as
\begin{eqnarray}
&&\vev{\phi (p_1) \cdots \phi (p_{n})}_{S (\Lambda)} 
\cdot (2 \pi)^4 \delta^{(4)}
(p_1 + \cdots + p_{n})\nonumber\\
&& \quad\equiv \frac{1}{Z}
\int [d\phi] \, \phi (p_1) \cdots \phi (p_{n})\, \e^{S (\Lambda)}
\end{eqnarray}
where
\begin{equation}
Z \equiv \int [d\phi] \,\e^{S(\Lambda)}
\end{equation}
Similarly, the correlation functions of a composite operator $\Op (p)$
is defined by
\begin{eqnarray}
&&\vev{\Op (p) \phi (p_1) \cdots \phi (p_n)}_{S(\Lambda)} \cdot (2
\pi)^4 \delta^{(4)} (p+p_1+\cdots+p_n)\nonumber\\
&& \quad\equiv \frac{1}{Z} \int [d\phi] \Op (p) \phi (p_1) \cdots \phi
(p_n) \e^{S(\Lambda)}
\end{eqnarray}

As a preparation, we first prove 
\begin{equation}
\left(- \Lambda \frac{\partial}{\partial \Lambda} + \sum_{i=1}^n
\frac{\Delta (p_i/\Lambda)}{K(p_i/\Lambda)} \right)
\vev{\Op (p) \phi (p_1) \cdots \phi (p_n)}_{S(\Lambda)} = 0
\label{compositelambda}
\end{equation}
for any composite operator that satisfies the ERG differential
equation
\begin{equation}
- \Lambda \frac{\partial \mathcal{O} (p)}{\partial \Lambda}
= \mathcal{D} \cdot \mathcal{O} (p)
\label{scalarcomposite}
\end{equation}
where
\begin{equation}
\mathcal{D} \equiv \int_q \frac{\Delta (q/\Lambda)}{q^2 + m^2} \left(
\frac{\delta \Si}{\delta \phi (q)} \frac{\delta}{\delta \phi (-q)} +
\frac{1}{2} \frac{\delta^2}{\delta \phi (q) \delta \phi (-q)}\right)
\end{equation}
(This result was first shown in \cite{becchi}.)
(Proof) Using (\ref{scalar}) and (\ref{scalarcomposite}), we obtain
\begin{eqnarray}
&&- \Lambda \frac{\partial}{\partial \Lambda} \left( \frac{1}{Z} \int
  [d\phi]\, \Op (p) \phi (p_1) \cdots \phi (p_n)
  \e^{S(\Lambda)}\right) \nonumber\\
&=& \frac{1}{Z} \int [d\phi] \, \phi (p_1) \cdots \phi (p_n)
\left( \Lambda \frac{\partial \ln Z}{\partial \Lambda} \Op (p) \e^S -
\Lambda \frac{\partial}{\partial \Lambda} \left( \Op (p) \e^S \right)
\right)\nonumber\\ 
&=& \int_q \frac{\Delta (q/\Lambda)}{q^2 + m^2} \frac{1}{Z}
 \int [d\phi] \, \phi (p_1) \cdots \phi (p_n) \nonumber\\
&& \times \left[ \frac{1}{2}
\frac{\delta^2}{\delta \phi (-q) \delta \phi (q)} \left( \Op (p) \e^S
\right) + \frac{\delta}{\delta \phi (q)} \left( \phi (q)
  \frac{q^2+m^2}{K(q/\Lambda)} \Op (p) \e^S \right) \right]\nonumber\\
&=& - \sum_{i=1}^n \frac{\Delta (p_i/\Lambda)}{K(p_i/\Lambda)} \cdot
\frac{1}{Z} \int [d\phi]\, \Op (p) \phi (p_1) \cdots \phi (p_n) \e^S
\end{eqnarray}
which gives (\ref{compositelambda}).

From (\ref{compositelambda}), we find that
\begin{equation}
\prod_{i=1}^n \frac{1}{K(p_i/\Lambda)} \cdot \vev{\Op (p) \phi (p_1)
  \cdots \phi (p_n)}_{S(\Lambda)}
\end{equation}
is independent of the cutoff $\Lambda$.

A particularly important example of a composite operator is
\begin{equation}
\Op (p) = \frac{\delta \Si}{\delta \phi (-p)}
\end{equation}
We can show that $\delta \Si/\delta \phi (-p)$ satisfies
(\ref{scalarcomposite}) by differentiating the ERG differential
equation (\ref{scalar}) with respect to $\phi (-p)$.

Now, we are ready.  Let us first consider the two-point function.
Using the Feynman rules, we immediately find
\begin{equation}
\vev{\phi (p) \phi (-p)}_{S(\Lambda)} = \frac{K(p/\Lambda)}{p^2 + m^2}
+ \frac{K(p/\Lambda)}{p^2 + m^2} \vev{\frac{\delta \Si}{\delta \phi
    (-p)} \phi (-p)}_{S(\Lambda)}
\end{equation}
\begin{figure}[t]
\begin{center}
\epsfig{file=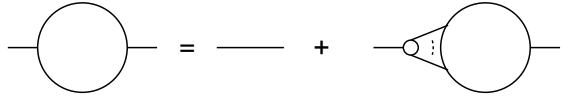}
\end{center}
\caption{The second graph has at least one interaction vertex.}
\end{figure}
The second term denotes the contribution of all the Feynman diagrams
with at least one interaction vertex.  (See Fig.~A1.)  This 
is an analogue of (\ref{firstrelation}).  Hence, we obtain
\begin{eqnarray}
&&\frac{1}{K(p/\Lambda)^2} \left[ \vev{\phi (p)
    \phi (-p)}_{S(\Lambda)} - \frac{K(p/\Lambda)}{p^2 + m^2}
\right]\nonumber\\
&& \qquad= \frac{1}{K(p/\Lambda)} \frac{1}{p^2 + m^2} \vev{\frac{\delta
    \Si}{\delta \phi (-p)} \phi 
  (-p)}_{S(\Lambda)}
\end{eqnarray}
Since $\delta \Si/\delta \phi (-p)$ is a composite operator, this is
independent of $\Lambda$. Thus,
\begin{equation}
\vev{\phi (p) \phi (-p)} \equiv \frac{1}{p^2 + m^2} +
\frac{1}{K(p/\Lambda)^2} \left[ \vev{\phi (p) 
    \phi (-p)}_{S(\Lambda)} - \frac{K(p/\Lambda)}{p^2 + m^2}
\right] 
\end{equation}
is independent of $\Lambda$.  This is the analogue of
(\ref{aa},\ref{psipsibar}).

For the higher-point correlation functions, there is no graph without
interaction vertices.  (See Fig.~A2.)
\begin{figure}[b]
\begin{center}
\epsfig{file=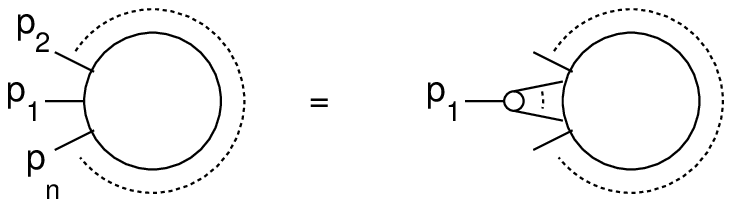}
\end{center}
\caption{There is at least one interaction vertex for the higher-point
  correlation functions.}
\end{figure}
Hence, we obtain
\begin{equation}
\vev{\phi (p_1) \cdots \phi (p_{n})}_{S (\Lambda)}
= \frac{K(p_1/\Lambda)}{p_1^2 + m^2} \vev{\frac{\delta \Si}{\delta \phi
    (-p_1)} \phi (p_2) \cdots \phi (p_{n})}_{S (\Lambda)}
\end{equation}
for $n > 1$.  This is an analogue of (\ref{secondrelation}).  Since
$\delta \Si/\delta \phi (-p_1)$ is a composite operator, we find
\begin{eqnarray}
&&\frac{1}{p_1^2 + m^2} \vev{\frac{\delta \Si}{\delta \phi
    (-p_1)} \phi (p_2) \cdots \phi (p_{n})}_{S (\Lambda)}
\prod_{i=2}^{n} \frac{1}{K(p_i/\Lambda)}\nonumber\\
&&\quad = \vev{\phi (p_1) \cdots \phi (p_{n})}_{S(\Lambda)}
\prod_{i=1}^{n} \frac{1}{K(p_i/\Lambda)}
\end{eqnarray}
is independent of $\Lambda$.  This is the analogue of (\ref{higher}).

An alternative derivation of the results given above can be found, for
example, in sect.~13 of \cite{wk} or in sect.~2 of \cite{IIS}, where
more general functional methods with external sources are used.

\section{Construction of $\Phi (-k)$}

The composite operator $\Phi (-k)$, introduced in section \ref{opid},
is given as the sum of two operators:
\begin{equation}
\Phi (-k) \equiv \Op_1 (-k) + \Op_2 (-k)
\end{equation}
where 
\begin{eqnarray}
  \Op_1 (-k) &\equiv& \int_p \Sp U (-p-k, p) \nonumber\\
  &&\quad \cdot \left\lbrace
    \ld{-p} S \cdot S \rd{p+k} + \ld{-p} S \rd{p+k} \right\rbrace\\
  \Op_2 (-k) &\equiv& e \int_p \Bigg[ - S \rd{p} \frac{K
    (p/\Lambda)}{K((p-k)/\Lambda)} \psi (p-k)\nonumber\\
  &&\qquad + \frac{K(p/\Lambda)}{K((p+k)/\Lambda)} \bar{\psi} (-p-k)
  \ld{-p} S \Bigg] 
\end{eqnarray}
Neither $\Op_1$ nor $\Op_2$ are composite operators by themselves.
(They do not satisfy $- \Lambda \frac{\partial}{\partial \Lambda}
\Op_i = \mathcal{D} \cdot \Op_i$.)  We wish to compute the correlation
functions for $\Op_1$, $\Op_2$ in order to show that $\Phi$ has the
desired correlation functions (\ref{Phicorr}).

The operator $\Op_2 (-k)$ is the change of the Wilson action
under the following linear change of variables:
\begin{eqnarray}
\delta \psi (p) &=& - e \frac{K(p/\Lambda)}{K((p-k)/\Lambda)} \psi
(p-k)\\
\delta \bar{\psi} (-p) &=& e \frac{K(p/\Lambda)}{K((p+k)/\Lambda)}
\bar{\psi} (-p-k)
\end{eqnarray}
Since the correlation functions are invariant under linear changes of
variables, the correlation functions of $\Op_2 (-k)$ are given by
\begin{eqnarray}
&&\vev{\Op_2 (-k) A_{\mu_1} (k_1) \cdots A_{\mu_M} (k_M) \psi (p_1)
  \cdots \psi (p_N) \bar{\psi} (-q_1) \cdots \bar{\psi} (-q_N)}_{S
  (\Lambda)}\nonumber\\
&& = - \sum_{i=1}^N \left[ \vev{A_{\mu_1} (k_1) \cdots \delta \psi (p_i)
    \cdots}_{S(\Lambda)} + \vev{A_{\mu_1} (k_1) \cdots \delta
    \bar{\psi} (-q_i) \cdots}_{S(\Lambda)} \right]
\end{eqnarray}
This gives
\begin{eqnarray}
  && \frac{1}{\prod_{i=1}^M K(k_i/\Lambda) \prod_{j=1}^N K(p_j/\Lambda)
    K(q_j/\Lambda)} \nonumber\\
  && \quad \cdot \vev{\Op_2 (-k) A_{\mu_1} (k_1) \cdots A_{\mu_M}
    (k_M) \psi (p_1) 
    \cdots \psi (p_N) \bar{\psi} (-q_1) \cdots \bar{\psi} (-q_N)}_{S
    (\Lambda)}\nonumber\\ 
  && = e \sum_{i=1}^N \left[ \vev{A_{\mu_1} (k_1) \cdots \psi (p_i - k)
      \cdots} - \vev{A_{\mu_1} (k_1) \cdots \bar{\psi} (-q_i - k) \cdots
    } \right] \label{O2}
\end{eqnarray}
for arbitrary $M, N$ except for $M=0, N=1$.  In the exceptional case
we obtain
\begin{eqnarray}
&&\frac{1}{K((q+k)/\Lambda) K(q/\Lambda)} \vev{\Op_2 (-k) \psi (q+k)
  \bar{\psi} (-q)}_{S(\Lambda)}\nonumber\\ && =
  \frac{e}{K(q/\Lambda)^2} \vev{\psi (q) \bar{\psi} (-q)}_{S
  (\Lambda)}\nonumber\\ &&\qquad\qquad - \frac{e}{K((q+k)/\Lambda)^2}
  \vev{\psi (q+k) \bar{\psi} (-q-k)}_{S (\Lambda)} \nonumber\\ && = \frac{U
  (-q-k, q)}{K((q+k)/\Lambda) K(q/\Lambda)}  \nonumber\\
&& \qquad + e \left( \vev{\psi (q) \bar{\psi} (-q)} - \vev{\psi
  (q+k) \bar{\psi} (-q-k)} \right) \label{O2exc}
\end{eqnarray}
where we used (\ref{psipsibar}) in the last step.  The first term on
the right-hand side is unwanted, and it is the role of $\Op_1$ to
remove it.

Let us compute the correlation functions for $\Op_1$.  For an arbitrary
insertion of gauge and $\psi, \bar{\psi}$ fields, we obtain
\begin{equation}
\int [dA d\psi d\bar{\psi}] \int_q
\frac{\overrightarrow{\delta}}{\delta \bar{\psi}_i (-q)} \left( \cdots
  \e^{S (\Lambda)} \right) \frac{\overleftarrow{\delta}}{\delta \psi_j (q+k)} 
  U_{ji} (-q-k,q) = 0
\end{equation}
This gives
\begin{eqnarray}
&& \vev{\Op_1 (-k) A_{\mu_1} (k_1) \cdots A_{\mu_M} (k_M) \psi (p_1)
    \cdots \psi (p_N) \bar{\psi} (-q_1) \cdots \bar{\psi}
    (-q_N)}_{S(\Lambda)}\nonumber\\ && - \sum_{i=1}^N \Bigg[
    \vev{\cdots U (-p_i, p_i - k) \ld{-p_i+k} S (\Lambda)
    \cdots}_{S(\Lambda)}\nonumber\\ && \quad\quad + \vev{ \cdots S
    (\Lambda) \rd{q_i + k} U (-q_i-k,q_i) \cdots}_{S(\Lambda)} \Bigg]
    = 0 
\end{eqnarray}
The case $M=0, N=1$ is exceptional because of the inhomogeneity of the
equations of motion:
\begin{eqnarray}
\vev{U (-q-k, q) \ld{-p} S (\Lambda) \cdot \bar{\psi}
    (-q)}_{S(\Lambda)} &=& - U (-q-k, q)\\
\vev{\psi (q+k) \cdot S (\Lambda) \rd{q+k} U (-q-k,q)}_{S (\Lambda)}
    &=& - U (-q-k,q)
\end{eqnarray}
Hence, we obtain
\begin{equation}
\vev{\Op_1 (-k) \psi (q+k) \bar{\psi} (-q)}_{S (\Lambda)} = -
U(-q-k,q)
\end{equation}
Thus, with (\ref{O2exc}) we obtain
\begin{eqnarray}
&&\vev{\Phi (-k) \psi (q+k) \bar{\psi} (-q)} \nonumber\\ &&\equiv
\frac{1}{K((q+k)/\Lambda) K(q/\Lambda)} \vev{ \left( \Op_1 + \Op_2
\right) (-k) \psi (q+k) \bar{\psi} (-q)}_{S (\Lambda)}\nonumber\\ &&=
e \left( \vev{\psi (q) \bar{\psi} (-q)} - \vev{\psi (q+k) \bar{\psi}
(-q-k)} \right)
\end{eqnarray}
For the cases other than $M=0, N=1$, the equations of motion are
homogeneous:
\begin{eqnarray}
\vev{\cdots U (-p_i, p_i - k) \ld{-p_i+k} S (\Lambda)
    \cdots}_{S(\Lambda)} &=& 0\\
\vev{ \cdots S
    (\Lambda) \rd{q_i + k} U (-q_i-k,q_i) \cdots}_{S(\Lambda)} &=& 0
\end{eqnarray}
(Note that only the connected parts are considered.)
Therefore, we obtain
\begin{equation}
\vev{\Op_1 (-k) A_{\mu_1} (k_1) \cdots \psi (p_1) \cdots \bar{\psi}
  (-q_1) \cdots }_{S(\Lambda)} = 0
\end{equation}
Hence, from (\ref{O2}) we obtain the desired result
\begin{eqnarray}
&& \vev{\Phi (-k) A_{\mu_1} (k_1) \cdots \psi (p_1) \cdots \bar{\psi}
    (-q_1) \cdots}\nonumber\\
&& = e \sum_{i=1}^N \left[ \vev{A_{\mu_1} (k_1) \cdots \psi (p_i - k)
      \cdots} - \vev{A_{\mu_1} (k_1) \cdots \bar{\psi} (-q_i - k) \cdots
    } \right]
\end{eqnarray}


\section*{References}

\begin{thebibliography}{99}
\bibitem{wk} Wilson K and Kogut J 1974 \textit{Phys. Rept.}
  \textbf{C12} 75
\bibitem{joe} Polchinski J 1983 \textit{Nucl. Phys.} \textbf{B231} 269
\bibitem{becchi} Becchi C 1991 On the construction of renormalized
  gauge theories using renormalization group techniques (Parma
  lectures) \textit{Preprint} \texttt{hep-th/9607188}
\bibitem{warr} Warr B 1988 \textit{Ann. Phys.} \textbf{183} 1; Warr B
  1988 \textit{Ann. Phys.} \textbf{183} 59
\bibitem{ellwanger} Ellwanger U 1994 \textit{Phys. Lett.}
  \textbf{B335} 364
\bibitem{bonini} Bonini M, D'Attanasio M and Marchesini G 1994
  \textit{Nucl. Phys.} \textbf{B418} 81
\bibitem{integral} Sonoda H 2003 \textit{Phys. Rev.} \textbf{D67}
  065011
\bibitem{erg} Sonoda H 2007 \textit{J. Phys.} \textbf{A40} 5733
\bibitem{scalar} Reuter M and Wetterich C 1994 \textit{Nucl. Phys.}
  \textbf{B427} 291
\bibitem{fw} Freire F and
  Wetterich C 1996 \textit{Phys. Lett.} \textbf{B380} 337 
\bibitem{arnone} Arnone S, Morris T and Rosten O 2005 \textit{JHEP}
  \textbf{0510} 115
\bibitem{rosten} Morris T and Rosten O 2006 \textit{J. Phys.}
  \textbf{A39} 11657
\bibitem{md} Morris T and D'Attanasio M 1996 \textit{Phys. Lett.}
  \textbf{B378} 213
\bibitem{IIS} Igarashi Y, Itoh K and Sonoda H 
  Quantum master equation for QED in ERG\\
  \textit{Preprint} \texttt{arXiv:0704.2349} 
\end{thebibliography}
\end{document}